# Bulk electronic properties of diamond

Christoph E. Nebel

Fraunhofer-Institute for Applied Solid State Physics (IAF), Tullastrasse 72, 79108 Freiburg, Germany

christoph.nebel@iaf.fraunhofer.de

## Abstract

A review of electronic properties of insulating-, boron- and phosphorus-doped diamond is given. The main goal is, to show data in a wider context, to reveal trends and limitations with respect to carrier mobilities, conductivities, p- and n-type doping. Undoped diamond is an insulator with conductivities significantly smaller than $10^{-17}$ $\Omega$cm at room temperature. Mostly, these insulating films show conductivity activation energies of 1.7 eV, an indication that small amounts of substitutional nitrogen dominates the Fermi-level. The electron and hole mobilities are very high (> 20.000 cm$^2$/Vs at T = 80 K) and are limited by acoustic phonon scattering. By phosphorus and boron doping diamond becomes semiconducting with an n-type donor activation energy (phosphorus) of 600 meV and an acceptor activation energy (boron) of 370 meV. Both dopands are hydrogen-like in nature. Due to the deep donor levels conductivity at room temperature is limited, electronic application will therefore be high-temperature devices. Both, electrons and holes show nearly the same mobilities which is promising for bi-polar applications, however, due to the very different doping levels low temperature devices will be governed by holes with increasing contribution of electrons towards higher temperatures. The effect of temperature on the carrier activation from donors and acceptors, on carrier scattering as well as on transport via hopping is shown in combination with some basic physical models and descriptions. The results confirm the superior electronic properties of diamond which makes it very promising for future electronic applications.

# Content

## 1. Introduction

Diamond is attracting more and more attention as it is becomes a mature wide-bandgap semiconductor material. The basic properties of diamond have been characterized over a long period of time in detail and some of the most important parameter are summarized in Table 1 in comparison to Si, GaN and 4H-SiC [1]. Several parameters of diamond are unique like high electron and hole mobilities both of about 2000 cm$^2$/Vs, thermal conductivity of 25 W/cmK, electric breakdown field of 10$^7$ V/cm, hardness of 10$^4$ kg/mm$^2$, and Debye temperature of 1860 K. It is therefore very attractive for a variety of applications. Due to the discovery of different methods for the production of diamond by man ("man-made diamond")

Diamond is nowadays available in single-, poly-, nano- and ultra-nano-crystalline structure. To fabricate diamond artificially, the high pressure high temperature (HPHT) technique [2], plasma-discharge-stimulated chemical vapor deposition methods [3-5] and the hot-filament technique [6] have been optimized during the last 60 years.

Most promising are optical applications (e.g. infrared windows, lenses, ATR units, X-ray windows), thermal applications (heatspreaders, laser submounts, X-ray targets), mechanical applications (cutting tools, scalpells, knives, length gauge tips, wear resistant components e.g. for textile machines, insert for dresser tools), electro-chemical applications (electrodes, electro-chemical detectors, bio-chemical sensors), radiation sensors (ionizing radiation detectors/dosimeters, fluorescence beam monitors), biolabels and drug delivery components (diamond nanoparticles, see also chapter 17). Based on single defects centers in diamond like the famous nitrogen vacancy center [7] quantum metrology is a promising field as well as quantum communication and computing (see also chapter 18). Both can emerge in the near future. Due to the progress of diamond doping, growth and technology [8,9] electronic applications in power devices are emerging [10]. Here, the missing single-crystalline diamond wafer is a major problem as this is a basic requirement for device fabrication in semiconductor production facilities. However, worldwide increasing activities with respect to hetero-epitaxial growth of single-crystalline diamond on Iridium will help to overcome this problem sooner or later [11-13].

The electronic properties of diamond are determined by the band structure, phonons, intrinsic and extrinsic defects as well as by dopant atoms. Among natural diamond, boron doped crystals (named IIb) are available and have been labeled "semiconducting diamond" [14]. Boron acts as acceptor, generating hole-conductivity with an activation energy of 360 meV. Synthetic high pressure high temperature diamond ("HPHT") can contain up to 1000 ppm of boron. Man-made diamond grown by plasma-enhanced CVD-technique can be boron doped in the regime $10^{16}$ cm$^{-3}$ to $10^{22}$ cm$^{-3}$ by adding gaseous compounds of boron to the plasma [15]. Nitrogen is the dominant extrinsic impurity in diamond [16]. Isolated substitutional nitrogen (P1 center) acts as a deep-donor with an ionization energy of 1.7 eV. In 1997, phosphorus doped CVD-diamond has been grown for the first time by adding phosphine (PH$_3$) to the plasma [8]. The incorporation of phosphorus is varied typically in the range $10^{16}$ to $10^{20}$ cm$^{-3}$[17]. Spectrally resolved photoconductivity experiments reveal an optical excitation energy of phosphorus of 0.56 eV with a hydrogen like electronic structure [18].

## 2.   Properties of insulating diamond

In general, no truly intrinsic diamond with a carrier activation energy of about 2.5 eV as expected for the large bandgap of 5.47 eV has been detected up to now. Estimating the density of free electrons n at 300 K (k$_B$T = 25 meV) in the conduction-band by:

$$n = N_c \exp\left[-\frac{E_c - E_F}{k_B T}\right] \quad \text{Eq. 1}$$

where the effective density of states of the conduction band, N$_c$, is assumed to be $\cong 6 \times 10^{19}$ cm$^{-3}$ and E$_c$ –E$_F$ = 2.5 eV. This results in an effectively zero ($10^{-19}$ 1/m$^3$) carrier density situation. The electronic properties of diamond will therefore be governed by dopants, defects, extrinsic contaminants and surface effects even if they are "ultra-low" in density.

A comparison of dark conductivities of very pure diamond (Type IIa) with man-made HPHT diamond of Type Ib (containing nitrogen) [19,20] and with typical high-quality poly-crystalline PECVD diamond layers [21] is shown in Fig. 1. The dark conductivity of type Ib diamonds show a thermal activation energy of 1.7 eV which is in agreement with the photoexcitation threshold of the P1 center (substitutional nitrogen) of 1.7 eV [22,23] as shown in Fig.2 where the absorption coefficients as function of photon energies as detected on typical type Ib (nitrogen containing HPHT), type IIa (ultra-pure diamond), type IIb (natural boron doped diamond) and intrinsic polycrystalline CVD diamond films (PCD) are shown. Based on these data it is reasonable to assume that the dark conductivity in Ib-diamond is n-type. The conductivity of type IIa diamond is about three orders of magnitude lower than type Ia diamond, but shows the same dark conductivity activation energy of 1.7 eV as Ib diamond. The purity of IIa diamond is reflected by the very low absorption coefficient in the band-gap regime. We assume that the activation energy of 1.7 eV in type IIa arises from a small amount of substitutional nitrogen atoms, which pin the Fermi level in the band-gap 1.7 eV below the conduction band minimum and cause the n-type nature. It is, however, not possible to generalize this conclusion. Redfield [24] for example detected by photo-Hall experiments in several highly resistive layers (comparable to type IIa diamond) a positive photo-Hall-voltage which indicates that some intrinsic diamonds can be p-type. At the end it will depend on the extrinsic contamination density of diamond which is either nitrogen or boron or both.

Fig. 1 shows also typical dark conductivities of high-quality intrinsic CVD-polycrystalline diamond layers [21]. No singly activated behavior can be observed. In the temperature regime 500 to 650 K, the activation energy $E_{act}$ is about 1 eV decreasing towards lower temperatures. The transport properties of polycrystalline diamond are dominated by conduction in the grains and in grain-boundaries which can be decorated with graphitic states [25,26]. The presence of sp2 states at grain boundaries is causing the continuous absorption spectra shown in Fig.2 and has been discussed in detail in the literature [25-27].

Hole mobility, drift velocity and effective mass

Mobilities of charge carriers in semiconductors are usually measured by the Hall effect technique. This method can, however, only been applied on conductive semiconductors. For highly resistive layers time-of-flight (ToF) experiments can be applied [28,29]. It is based on a pulsed photo-excitation where electron-hole pairs are generated for example by a pulsed laser illumination. Using strongly absorbing light the carriers will be generated close to a contact and can be separated by an external applied electric field. Dependent on the electric field and the generation of carriers, either electrons or holes are propagating and their drift velocity and related mobility can be detected as function of temperature and applied electric field.

Fig. 3 summarizes published drift [29-31] and Hall mobilities [32,33] of holes, as detected on highly insulating and natural semiconducting diamond films. The data deduced by these different research teams and at very different times show the same tendency. The quality of diamonds (defects, extrinsic contaminations, dopands) could be slightly different, thereby causing variations as shown. The main features of the shown mobilities are:

1) Below 300 K, the hole drift mobility exhibits the typical $T^{-3/2}$ temperature dependence expected for acoustic phonon scattering. For higher temperatures the slopes become steeper showing a $T^{-s}$ dependence with s = 2.7 – 2.9 [25] indicating the onset of intervalley phonon scattering [22].

2) As shown in Fig. 3, for intrinsic diamond the hole mobility at 290 K is in the range 1500 – 2300 cm$^2$/Vs. With increasing electric field strength the drift velocity of holes exhibits a sublinear increase [30] as shown in Fig.4 which is characteristic for covalent semiconductors [37]. For T ≤ 300 K an anisotropic drift velocity is observed, with $v_{d<100>} \geq v_{d<110>}$. Reggiani et al. [34] attribute this result to the warped and nonparabolic features of the valence band. A saturated hole drift velocity of 1.1 (±0.1)x10$^7$ cm/s is detected for electric fields ≥ 25 kV/cm and temperatures T < 300 K.

The limiting drift velocity in a semiconductor with dominant carrier scattering by optical phonons is given by [38]:

$$v_{sat} = \left[ \frac{8E_{opt}}{3\pi m^*} \right]^{1/2} \quad \text{Eq. 2}$$

where $E_{opt}$ is the energy of the optical phonon and m* the density of states effective hole mass. Reggiani et al. [34] calculate from these considerations the heavy (hh) and light (lh) hole masses to be $m_{hh}$ =1.1 $m_o$ and $m_{lh}$ = 0.3 $m_o$, respectively ($m_o$ is the electron rest mass). A detailed discussion will be given below.

Electron mobility, drift velocity and effective mass

To characterize electron mobilities time-of-flight and photo-Hall experiments have been applied on the same samples as shown for holes in Fig. 3. The results are summarized in Fig. 5 [24,31,33,39]. The main features of electron transport have been discussed by Nava et al. [39] and Isberg et al. [31] and can be summarized as follows:

1)      As shown in Fig. 5 the electron mobility at 300 K is in the regime 1900 to 2500 cm$^2$/Vs. As for holes, at temperatures < 300 K a typical T$^{-3/2}$ dependence is found (acoustic phonon scattering) with a highest mobility at 92 K of 19000 cm$^2$/Vs. Above 300 K the slopes becomes steeper which has been be attributed to phonon induced intervalley scattering.

The electron drift velocity versus electric field (see Fig.6) shows also an anisotropy with $v_{<100>}$ larger than $v_{<110>}$. This anisotropy is characteristic for a multi-valley band structure with ellipsoidal constant energy surfaces elongated along equivalent <100> directions. The saturation drift velocity is 1.5(±0.1)x10$^7$ cm/s for all temperatures below 300 K.

Based on drift mobility data, Nava et al. [39] have estimated the transverse (or $m_\perp$) and longitudinal (or $m_\parallel$) effective mass of electrons in diamond to $m_t = 0.36\ m_o$ and $m_l = 1.4\ m_o$, which is in reasonable agreement with results from band structure calculations [40] from Willatzen and Cardona who applied linear muffin-tin-orbital and kp calculations of the effective masses and band structure of semiconducting diamond. They deduce $m_t$ (or $m_\perp$) = $0.34\ m_o$ and $m_l$ (or $m_\parallel$) = $1.5\ m_o$.

### 3. Properties of doped diamond

At the very early stage of electronic characterization by Austin and Wolfe in 1956 [41] boron doped diamond named "natural semiconducting diamond" has been investigated by conductivity experiments which revealed an activation energy of 0.37 eV. At that time aluminum was assumed to be the doping element, which was later corrected to boron. Collins and Lightowlers [42] in 1968 used photothermal ionization spectroscopy to reveal the hydrogen-like nature of the acceptor spectrum with an ionization energy threshold of 0.373 eV.

Nearly thirty years later, in 1997, phosphorus doping of diamond has been demonstrated for the first time generating n-type properties [8,44]. Again, Hall-effect [8,43] and photoionization spectroscopy [18, 44] was applied to identify the thermal activation energy of phosphorus to be 0.6 eV as well as its hydrogen-like nature [45,46]. Fig. 7 shows a comparison of phosphorus (n-type) and boron doped (p-type) diamond containing $5 \times 10^{17}$ cm$^{-3}$ boron and $8 \times 10^{17}$ cm$^{-3}$ phosphorus (data from Ref. 17). Due to the difference in activation energies of 0.37 eV for boron (with respect to the valence band minimum) and 0.6 eV for phosphorus (with respect to the conduction band minimum) the carrier concentration difference at room temperature is 5 orders of magnitude.

Electronic properties of p-type diamond

For a non-degenerate p-type semiconductor (equivalently for n-type material) containing an acceptor density $N_A$ and a density of compensating donors, $N_D$, the density-of-holes p at a given temperature T in the valence band can be calculated using [47]:

$$\frac{p(p + N_D)}{(N_A - N_D - p)} = \frac{2}{\beta} \left\{ \frac{2\pi m^* k_B T}{h^2} \right\}^{3/2} \exp\left[ -\frac{E_A}{k_B T} \right] \qquad \text{(Eq. 3)}$$

where β is the spin degeneracy of acceptors, m* is the effective density of state hole mass, $E_A$ is the acceptor ionization energy, $k_B$ is the Boltzmann constant and T the temperature.

Hole and electron effective density of state masses

To use Eq. 3 we need to know the effective density of state hole mass which can be calculated from the band-structure of diamond. The band-structure of diamond has been discussed by Willatzen and Cardona [40] in 1994 using ab initio calculations based on the linear muffin-tin-orbital method in local-density approximation which is well known to be one of the most accurate method to determine the band structure, and so the effective masses, in semiconductors. They show that the three uppermost valence-bands have their maximum localized at the Γ point of the Brillouin zone. The two upper bands named heavy-hole (hh) and ligh- hole (lh) bands are degenerated at the Γ point and energetically positioned at a higher energy than the third hole band named split-off band (so). They calculate 13 meV energy separation which is close to the experimentally detected value of 6 meV [30,34,48]. The three valence bands are considered to be parabolic. For temperatures higher than 70 K (> 6 meV) the three bands can be considered as degenerate at the Γ point of the Brillouin zone.

Hole effective masses

The values for the heavy hole (hh), light hole (lh) and the split-off (so) masses from Willatzen and Cardona are [40]:

$$m_{hh}^{*100} = 0.427 \, m_o$$

$$m_{hh}^{*110} = 0.69 \, m_o$$

$$m_{lh}^{*100} = 0.366 \, m_o$$

$$m_{lh}^{*110} = 0.276 \, m_o$$

$$m_{so}^{*} = 0.394 \, m_o$$

where $m_o$ is the electron rest-mass.

The corresponding values for the density of state mass of the heavy hole (hh) is given by [49]:

$$m_{hh}^* = \{(m_{hh}^{*110})^2 \ m_{hh}^{*100}\}^{1/3} = 0.588 \ m_o$$

The density of state light hole (lh) mass is given by:

$$m_{lh}^* = \{(m_{lh}^{*110})^2 \ m_{lh}^{*100}\}^{1/3} = 0.303 \ m_o$$

The total density of valence band state effective mass m* can be calculated by [49]:

$$m^* = \left\{ m_{lh}^{*\frac{3}{2}} + m_{hh}^{*\frac{3}{2}} + m_{so}^{*\frac{3}{2}} \right\}^{\frac{2}{3}} = 0.908 \ m_o \qquad \text{(Eq. 4)}$$

Free hole density for p << N$_A$ and N$_D$

At high temperatures, p approaches the saturation value N$_A$-N$_D$ and at low temperatures, where p << N$_A$, N$_D$, Eq. 3 results in [51]:

$$p \approx \left( \frac{N_A}{N_D} - 1 \right) \frac{2}{\beta} \left\{ \frac{2\pi m^* k_B T}{h^2} \right\}^{3/2} \exp\left[ -\frac{E_A}{k_B T} \right] \qquad \text{(Eq. 5)}$$

When an impurity level is created by splitting of states from the conduction or valence band with multiple or degenerate extrema, the impurity level spin degeneracy β will be larger than two. For acceptor levels introduced by group III impurities in Si and Ge, β = 4, because the heavy hole and light hole bands are degenerate at $\boldsymbol{k}$ = 0 and the spin-orbit splitting energy $\varDelta$ = 44 meV in Si and 295 meV in Ge is much larger than k$_B$T, so that in general the split-off band is well separated and thermally occupied only to a negligible degree [47,52]. In diamond however, spin-orbit coupling is week, $\varDelta$ = 6 meV [30,34,48] so that a threefold band degeneracy with β = 6 at temperatures above about 70 K can be expected. Below 70 K, a twofold band degeneracy with β = 4 is a better approximation.

The concentration p of holes is generally obtained from Hall effect data as:

$$p = \frac{r_H}{\text{Re}} \qquad \text{(Eq. 6)}$$

where R is the Hall coefficient, r$_H$ is the Hall factor and e the elementary charge. For acoustic deformation potential scattering, $r_H = 1.18$ while for ionized impurity scattering $r_H = 1.93$ [53]. Most Hall data in the literature have been analyzed by assuming $\beta = 2$ and $r_H = 1.18$. The discrepancy with the correct values discussed above makes relatively little difference for a quantitative analysis since effectively $(m^*)^{3/2}/\beta$ in Eq. 3 and 5 is used as an adjustable parameter to obtain the best fits to the experimental data.

Fig. 8 shows a plot of the hole density versus reciprocal temperature for a natural type IIb diamond [35], showing the predicted features given by Eq. 3 and 5. The evaluation results in a boron density of $N_A \cong 8 \times 10^{16}$ cm$^{-3}$, a compensating donor density of $N_D \cong 10^{15}$ cm$^{-3}$ and a boron ionization energy of $E_A \cong 368$ meV which is in good agreement with the photo-ionization energy of 373 meV [42].

Free hole density for p >> $N_D$

If the acceptor density $N_A$ is significantly larger than the compensating donor density $N_D$ ("nearly uncompensated semiconductor") and for the case that the free hole density in the valence band p >> $N_D$ (which is the case at sufficiently high temperatures), the temperature dependent variation of p is given by [47]:

$$p(T) \cong \left\{ \frac{N_V}{\beta} (N_A - N_D) \right\}^{1/2} \exp\left( -\frac{E_A}{2k_B T} \right) \qquad \text{(Eq. 7)}$$

where $N_V$ is the effective density of valence band states, which varies with temperature as $N_V = N_{Vo} T^{3/2}$.

Fig. 9 shows typical hole densities measured as a function of temperature by Hall experiments on samples with different boron doping densities variing from $6.4 \times 10^{16}$ cm$^{-3}$ to $1.3 \times 10^{18}$ cm$^{-3}$ to $10^{20}$ cm$^{-3}$ (data from Teraji et al [54]). The resistivity of the highest doped sample reaches 0.05 $\Omega$cm at 650 K which is about comparable to values of Silicon with a doping density of ca. $10^{18}$cm$^{-3}$ at the same temperature. The hole concentration shows thermally activated behavior and is increasing with increasing doping density. From these data the boron acceptor activation energy $E_A$ is about 0.36 eV for doping densities below $10^{18}$ cm$^{-3}$. At higher doping densities, $E_A$ decreases gradually and becomes 0.29eV at ca. $10^{20}$ cm$^{-3}$. It is interesting to note that the activation energy of doped natural conducting diamond is slightly larger 0.368 eV [35] than these CVD layers with 0.36 eV activation energy. This may arise by energy level broadening of the excited states of boron due to constrain which is larger in CVD than in natural diamonds [55].

Hall mobilities measured on the same samples are shown in Fig. 10 as a function of temperature. Highest hole mobilities are detected on the sample with the lowest doping density. For a doping density of $6.4 \times 10^{16}$ cm$^{-3}$ it rises from 1620 cm$^2$/Vs at 290K to 2750 cm$^2$/Vs at 215 K. The Hall mobility μ is decreasing towards high temperature due to scattering with optical phonons while at lower temperature scattering with ionized impurities (compensating defects, ionized acceptors) is dominant.

A comparison with hole mobility data measured on undoped diamond film with photoexcitation techniques shows reasonable agreement. The low temperature regime T < 280 K is dominated by scattering of holes with ionized impurities in doped films while this is negligible in undoped diamond. Here the dominant scattering mechanism is acoustic phonon scattering (T$^{-3/2}$). For higher temperatures the slope becomes steeper indicating the onset of intervalley phonon scattering in undoped diamond [29,49]. At temperatures > 500 K the scattering with optical phonons becomes dominant in diamond.

Hole conductivity

The typical dependence of the conductivity on reciprocal temperature for different boron concentrations is shown in Fig. 11 [19]. The measured electrical conductivity can be described phenomenologically by a combination of various transport mechanisms:

$$\sigma = \sum^{i} \sigma_i \quad \text{(Eq. 8)}$$

In the high temperature regime, holes are thermally excited from acceptor states into the valence band. In this regime, σ is thermally activated with E$_{act}$ = 370 meV, the acceptor ionization energy. At lower temperatures or higher doping densities, holes propagate via variable range hopping first discussed by Mott [56] where:

$$\sigma = \sigma_{hop} \exp\left\{ -\left( \frac{T_o}{T} \right)^{1/4} \right\} \quad \text{(Eq. 9)}$$

with T$_o$ given by:

$$T_o = \frac{18.1}{k_B D(E_F) \alpha^3} \quad \text{(Eq. 10)}$$

D(E$_F$) is the density-of-states at the Fermi level E$_F$ and α is the effective Bohr radius of boron acceptor levels. The conductivity pre-factor σ$_{hop}$ can be calculated by:

$$\sigma_{hop} = e^2 D(E_F) R^2 \nu_{ph} \quad \text{(Eq. 11)}$$

where R is the average hopping distance and $\nu_{ph}$ the corresponding exchange frequency (Raman frequency $4 \times 10^{13}$ 1/s). Based on these equations, reasonable fits have been achieved for the low temperature data (see for example Ref. [19,57,58]) which result in a Bohr radius $\alpha$ of about 6-8 Å. At room temperature, hopping dominates transport at doping levels $\geq 5 \times 10^{19}$ cm$^{-3}$. Here the activation energy drops rapidly towards zero (Fig. 12), in conjunction with a decrease of the resistivity over three orders of magnitude (Fig. 13). Also shown in Fig. 12 is the evolution predicted by the model of Pearson and Bardeen [59]. At acceptor densities exceeding $3 \times 10^{20}$ cm$^{-3}$, a metal-insulator transition takes place [14,19]. Such heavily doped samples show an increasing resistivity with increasing temperature. The metallic resistivity in boron doped diamond is about $10^{-3}$ to $10^{-2}$ $\Omega$cm.

Electronic properties of n-type diamond

The free electron density

The variation of the electron density as function of temperature as detected by Hall effect measurements on (111)-oriented CVD diamond films with phosphorus densities $2.3 \times 10^{16}$ cm$^{-3}$, $8.9 \times 10^{16}$ cm$^{-3}$ and $2 \times 10^{18}$ cm$^{-3}$ are shown in Fig. 14 [60]. Between 300 and 870 K these films show negative Hall coefficients, which indicate n-type conduction. The free electron density in the conduction band is exponentially activated with 0.57 eV in the temperature regime 300 to 500 K. To evaluate most important parameters Eq. 12 has been used where $N_D$ is the phosphorus density, $N_A$ is the compensating acceptor density, n is the free electron density in the conduction band, $N_C$ is the effective conduction band density of states, g is the degeneration factor of donors, $E_D$ is the activation energy of donors, $k_B$ is the Boltzmann constant and T is the temperature.

$$\frac{n(n+N_A)}{N_D - N_A - n} = \frac{N_C}{g} exp\left[-\frac{E_D}{k_B T}\right] \qquad \text{Eq. 12}$$

For a phosphorus-doped films with a P concentration of $7 \times 10^{16}$ cm$^{-3}$ best fits to the thermal variation of electrons have been achieved with an activation energy $E_D$ of 0.57 eV, a doping density $N_D$ of $6.8 \times 10^{16}$ cm$^{-3}$ and a compensating defect density $N_A$ of $8.8 \times 10^{15}$ cm$^{-3}$. This results in a compensation ratio $N_A/N_D$ of 0.13. With decreasing P concentration below $10^{16}$ cm$^{-3}$, the phosphorus doped films become fully compensated and highly resistive. It indicates that the concentration of compensating defects $N_A$ in (111)-oriented CVD diamond is of the order of $10^{15}$ cm$^{-3}$.

Electron effective mass

For the discussion of electron data, we need to consider six equivalent minima of the conduction band which are located in the Brillioun zone at $2\pi/a$(0.742, 0, 0) close to the X point. We have to consider two effective masses for the principal directions $m_\parallel$ in (100) direction (longitudinal) and $m_\perp$ (transversal) in (010) direction. According to Ref. 40 the effective masses are $m_\parallel = 1.5\ m_o$ and $m_\perp = 0.341\ m_o$.

Pernot and coworkers [49,50] deduce as effective Hall mobility mass for electrons:

$$m_H^* = \frac{2m_\parallel + m_\perp}{2 + m_\parallel/m_\perp} \qquad \text{Eq. 13}$$

Electron Hall mobilities

Figure 15 summarizes typical Hall mobilities of phosphorus doped films as a function of temperature [60]. The Hall mobility depends on the doping density. At 300 K it is decreasing from 660 cm$^2$/Vs for a phosphorus density of $7\times10^{16}$ cm$^{-3}$, to 330 cm$^2$/Vs for $10^{18}$ cm$^{-3}$, to 180 cm$^2$/Vs for $3\times10^{18}$ cm$^{-3}$. A detailed discussion of scattering mechanisms by Pernot and co-workers [51,60] shows that the mobilities at RT are dominated by intravalley acoustic scattering. This arises by the low density of compensating centers making ionized and neutral impurity scattering not very effective. At higher temperatures T > 700 K intervalley scattering becomes dominant.

Fig. 15 shows also electron mobilities detected by Time-of-flight and Photo-Hall experiments on insulating diamond (see Fig. 5). The T$^{-3/2}$ temperature dependence at low temperature has been attributed to acoustical phonon scattering and seems to dominate the mobility in the temperature regime between 80 and 400 K. Pernot [50] concludes, that the difference in absolute values cannot be described with a classical model using only one deformation potential related to the conduction band of diamond. He suggests a major difference between experiments performed in thermodynamic equilibrium (Hall experiments) and experiments which are based on photo-excitation.

<u>n-type conductivity</u>

Finally, typical temperature dependent conductivities σ(T) of the n-type diamond films are shown in Fig. 16 [61]. At lower doping densities ($3\times10^{18}$ cm$^{-3}$) the conductivity is single activated and can be described by:

$$\sigma(T) = \sigma_o exp^{-\frac{E_{act1}}{k_B T}} \qquad \text{Eq. 14}$$

where σ$_o$ is the pre-factor of the conductivity, E$_{act1}$ is the activation energy for band conduction, T is the temperature and k$_B$ is Boltzmann constant.

The pre-factor of conductivity is:

$$\sigma_o = q\, n_o\, \mu \qquad \text{Eq. 15}$$

where q is the elementary charge, n$_o$ is the density of free electrons in the conduction-band and μ is the mobility of electrons. This results in 0.56 eV activation energy and indicates thermal activation of electrons from the phosphorus doping level to the conduction band in reasonable agreement with Hall effect measurements which reveal 0.58 eV activation energy. The activation energies E$_{act1}$ measured at high temperatures (T > 500 K) are decreasing from 0.56 eV at doping levels $3\times10^{18}$ cm$^{-3}$ and $4\times10^{19}$ cm$^{-3}$ to about 0.46 eV at $10^{20}$ cm$^{-3}$ due to the temperature induced decrease of the electron mobility towards higher temperatures.

For higher doping densities σ(T) can be described by a combination of extended state transport (E$_{act1}$) and thermally stimulated hopping transport governed by thermally activated (E$_{act2}$) tunneling transitions (αR):

$$\sigma(T) = \sigma_0 exp^{(-\frac{E_{act1}}{kT})} + \sigma_1 exp^{(-2aR - \frac{E_{act2}}{kT})} \quad \text{Eq. 16}$$

The hopping transport of electrons moving thermally assisted from occupied to nearby unoccupied state is governed the transition probability P which is:

$$P(T) = \nu_r exp^{(-2aR - \frac{E_{act2}}{kT})} \quad \text{Eq. 17}$$

where $v_r$ is the Raman frequency $4 \times 10^{13}$ 1/s, $\alpha$ is the localization length of an electron in a donor state and R is the distance to the nearest unoccupied state which can also show an energy barrier of $E_{act2}$ between both states. The discussion of these data by Matusmoto and co-workers [61] reveal a Bor radius of phosphorus in diamond of 3 to 4 Å. This is about half the value of the localization length of holes in the boron level. Evaluating the slope at low temperatures reveal activation energies $E_{act2}$ of 22 meV for $4 \times 10^{19}$ cm$^{-3}$, of 37 meV for $7 \times 10^{19}$ cm$^{-3}$ and 48 meV for the highest doping level of $10^{20}$ cm$^{-3}$. This slight increase with increasing doping is attributed to an increase in Coulomb repulsion. Obviously, the phosphorus doping level ground state becomes energetically broader with increasing phosphorus doping. Due to a Coulomb repulsion effect a direct correlation with the measured energies is not possible.

## 4. Summary

Undoped diamond is an insulator with highest electron and hole mobilities. The conductivity at 300 K is significantly smaller than $10^{-17}$ $\Omega$cm showing typical conductivity activation energies of 1.7 eV. This is attributed to small amounts of substitutional nitrogen which dominates the Fermi-level. The electron and hole mobilities are very high (> 20.000 cm$^2$/Vs at T = 80 K) and are limited by acoustic phonon scattering. By phosphorus and boron doping diamond becomes semiconducting with an n-type donor activation energy (phosphorus) of 600 meV and an acceptor activation energy (boron) of 370 meV. Both dopands are hydrogen-like in nature. Due to the deep donor levels conductivity at room temperature is limited, electronic application will therefore be high-temperature devices. As both, electrons and holes show nearly the same mobilities true bipolar properties could be expected, however, due to the very different doping levels low temperature devices will be governed by holes with increasing contribution of electrons towards higher temperatures. The effect of the temperature on the carrier mobility can be split into two regimes: at low temperature acoustic phonon scattering limits the mobility while at high temperature intervalley phonon scattering seems to dominate the propagation. Effective masses of carriers have been introduced as well as the saturation velocity of electrons and holes. The effect of temperature on the carrier activation from donors and acceptors as well as on transport via hopping propagation has been shown in combination with some basic physical models and descriptions. The results confirm the superior electronic properties of diamond which makes it very promising for future electronic applications.

**Figure Captions:**

**Fig. 1**

Conductivity properties of natural type IIa, type Ib and of poly-crystalline CVD diamond. The very pure type IIa diamond shows the lowest conductivity which is activated with 1.7 eV [20]. Several Ib-synthetic HPHT diamonds also show 1.7 eV activation energies but they are about three orders of magnitude higher in conductivity [19]. Transport in polycrystalline CVD-diamond is activated with energies increasing with increasing temperature [21].

**Fig. 2**

Optical absorption coefficients of type IIa, type Ib and polycrystalline CVD diamond (PCD) as measured by transmission/reflectivity and photo-thermal deflection spectroscopy (PDS) experiments.

**Fig. 3**

Hole mobilities as a function of temperature in natural diamond. The data have been determined by Hall effect (Dean [32], Konorova [33]) and time-of-flight experiments (Reggiani [30], Isberg [29]). The continuous curve refers to theoretical calculations by Reggiani et al. [30].

**Fig. 4**

Hole drift velocity as a function of electric field measured at a temperature of 85 K in natural diamond [30]. An characteristic anisotropy $v_{<100>} > v_{<110>}$ is detected.

**Fig. 5**

Electron mobilities as a function of temperature in natural diamond. The data have been published by Redfield [24] (Hall effect), Konorova [33] (Hall effect), by Nava [39] and Isberg [29] (time-of-flight). The solid line refers to theoretical calculations of Nava et al [39].

**Fig. 6**

Electron drift velocity as function of electric field in natural diamond [39] detected at 85 K. The closed and open circles refer to experiments performed along <110> and <100> crystallographic directions.

**Fig. 7**

Tempereture dependence of carrier concentrations obtained for p-type boron doped and n-type phosphorus doped diamond films grown on Ib (001)-oriented substrates [17].

**Fig. 8**

Hole concentration in the valence band as function of reciprocal temperature [35]. The By calculating fits of Eq. ?? to the date the authors reveal an acceptor density is $8x10^{16}$ cm$^{-3}$ and the compensating donor density of about $10^{15}$ cm$^{-3}$.

**Fig. 9**

Hole concentration plotted as a function of inverse of temperature for B-doped diamond films grown using high-power MPCVD [55]. The hole concentration shows a thermally activated behavior. The boron acceptor activation energy $E_A$ is 0.36eV up to $N_A$ of ca. $10^{18}$cm$^{-3}$; above this density $E_A$ decreases gradually to become 0.29eV at $N_A$ of $10^{20}$cm$^{-3}$.

**Fig. 10**

Hall mobilities of boron doped CVD diamond as function of temperature are shown as detected by Hall-effect measurements and compared to mobilities measured on insulating diamond by photo-excitation techniques. High Hall mobilities of 1620 cm$^2$/Vs at 290 K and of 2750 cm$^2$/Vs at 215 K were detected on the $6.4x10^{16}$ cm$^{-3}$ doped layer. The Hall mobilities show a steeper variation as function of temperature than the data from undoped diamond.

**Fig. 11**

Conductivity data of boron doped diamond [19]. The acceptor density is indicated in the plot in units of cm$^{-3}$. The hopping regime at low temperature is followed by a thermally activated regime where $E_{act}$ = 370 meV. With increasing doping the onset of hopping shifts towards higher temperatures.

**Fig. 12**

Conductivity activation energies at 300 K as function of boron content. At about [B] > $3x10^{18}$ cm$^{-3}$ the activation energy starts to decrease and finally vanishes for [B] > $2x10^{20}$ cm$^{-3}$. The theoretical dependence according to the model of Pearson and Bardeen (solid curve) is given for comparison (solid line) [59].

**Fig. 13**

Variation of the resistivity of boron doped diamond at 300 K with boron doping levels. Below about $2 \times 10^{19}$ cm$^{-3}$, the decrease is inversely proportional to the acceptor density. In the range $2 \times 10^{19}$ to $3 \times 10^{20}$ cm$^{-3}$ hopping becomes dominant. For higher concentrations metallic conductivity is observed.

**Fig. 14**

Temperature dependence the carrier concentrations of P-doped diamond films with various P concentrations are shown [60]. The activation energy of these layers is about 0.57 eV. A compensation ratio of about 0.13 is detected for the film with $8.9 \times 10^{16}$ cm$^{-3}$ donors incorporated.

**Fig. 15**

Hall mobilities of differently phoshorus doped diamond films, grown by plasma enhance CVD [60] are shown in comparison to mobility data measured by phot-excitation in insulating layers. The Hall mobility at RT increases from 410 to 660 cm$^2$/Vs with decreasing P concentrations from $5 \times 10^{17}$ to $7 \times 10^{16}$ cm$^{-3}$. The Hall mobility for the P-doped film with $7 \times 10^{16}$ cm$^{-3}$ decreases with increasing temperature as $T^{-1.4}$ up to 450 K. This indicates acoustic phonon scattering. At temperatures above 450 K, the Hall mobility is proportional to $T^{-2.6}$ which is attributed to various scattering mechanisms such as intervalley scattering, as well as acoustic phonon scattering.

**Fig. 16**

Temperature dependence of the conductivity for P-doped n-type diamond films measured between 200 and 1000 K [61]. The doping densities are indicated. The properties are semiconducting with two transport regimes. At lower temperatures the transport is dominated by a thermally activated tunneling transport, whereas at higher temperatures the conductivity is dominated by free electrons in the conduction band.

| Properties/Materials | Diamond | Si | 4H-SiC | GaN |
|---|---:|---:|---:|---:|
| Bandgap (eV) | 5,47 | 1,12 | 3,26 | 3,5 |
| Electron-Mobility ($cm^2$/Vs) bei 300 K | 1.900 -2.300 | 1.500 | 900 | 1.250 |
| Hole Mobility ($cm^2$/Vs) bei 300 K | 1.500 - 2.300 | 600 | 100 | 200 |
| Dielectric Constant | 5,7 | 11,9 | 9,7 | 9,5 |
| Thermal Conductivity (W/cmK) | 25 | 1,48 | 4,9 | 1,3 |
| Electron-Saturation Velocity ($10^7$ cm/s) | 2,7 | 1 | 2,7 | 2,7 |
| Breakdown Field ($10^5$ V/cm) | 100 | 3 | 30 | 30 |
| Debye Temperature (K) | 1.860 | 645 | 1.200 | 608 |
| Hardness (kg/$mm^2$) | 10.000 | 1.000 | 4.000 | |
| Johnson's FoM | 81.000 | 1 | 278 | 215 |
| Baliga's FoM | 25.100 | 1 | 125 | 187 |
| Bipolar Power Switching Product | 1.426.711 | 1 | 748 | 560 |

Table 1

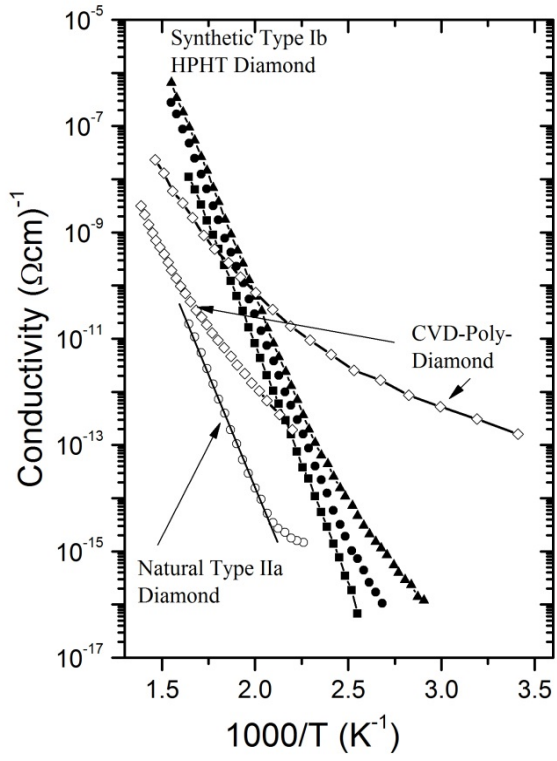

Fig.1

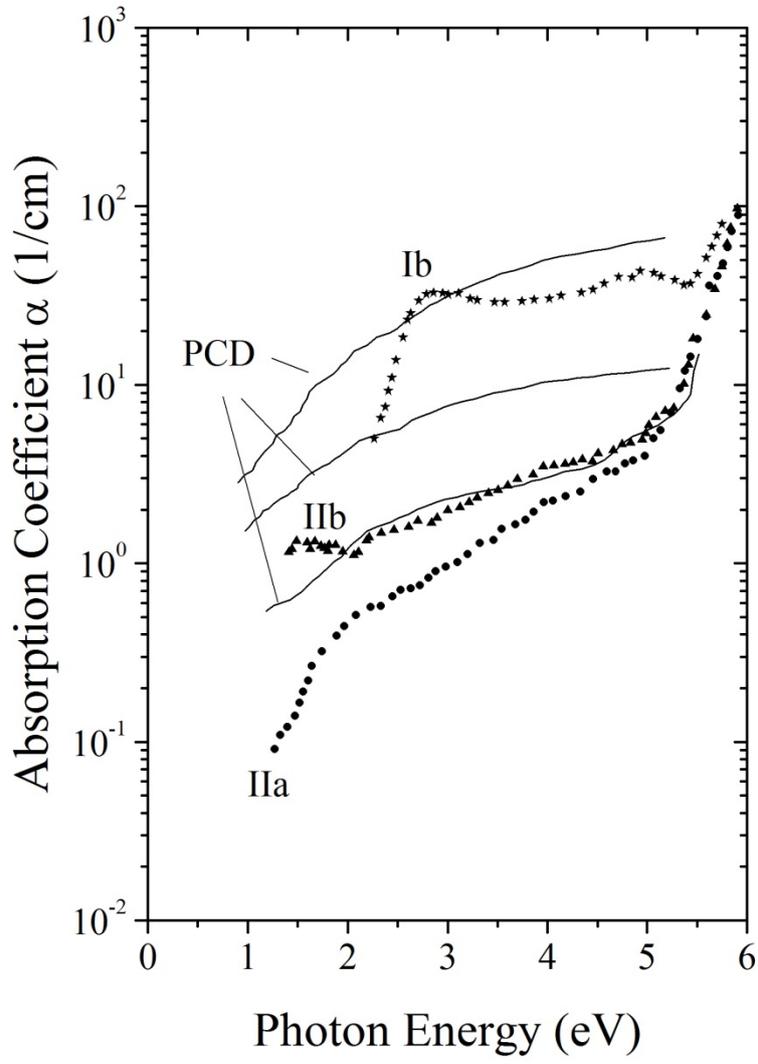

Fig. 2

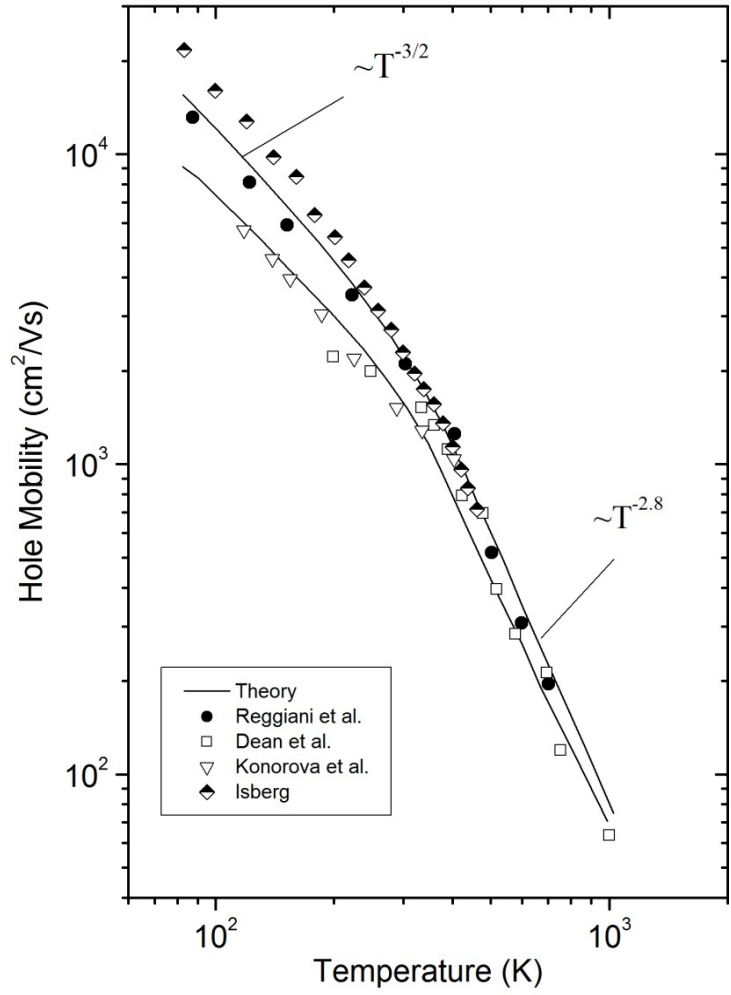

holemob.opj

Fig. 3

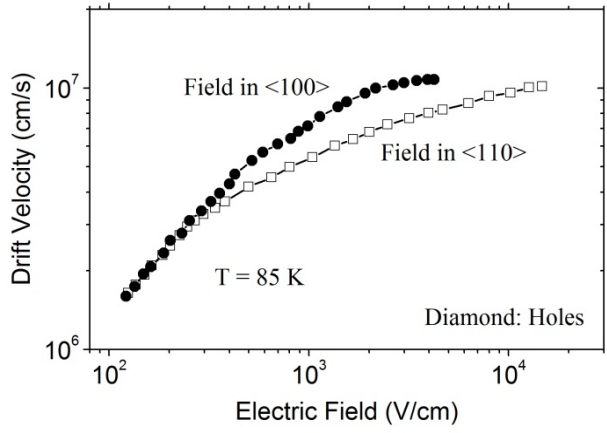

Fug. 4

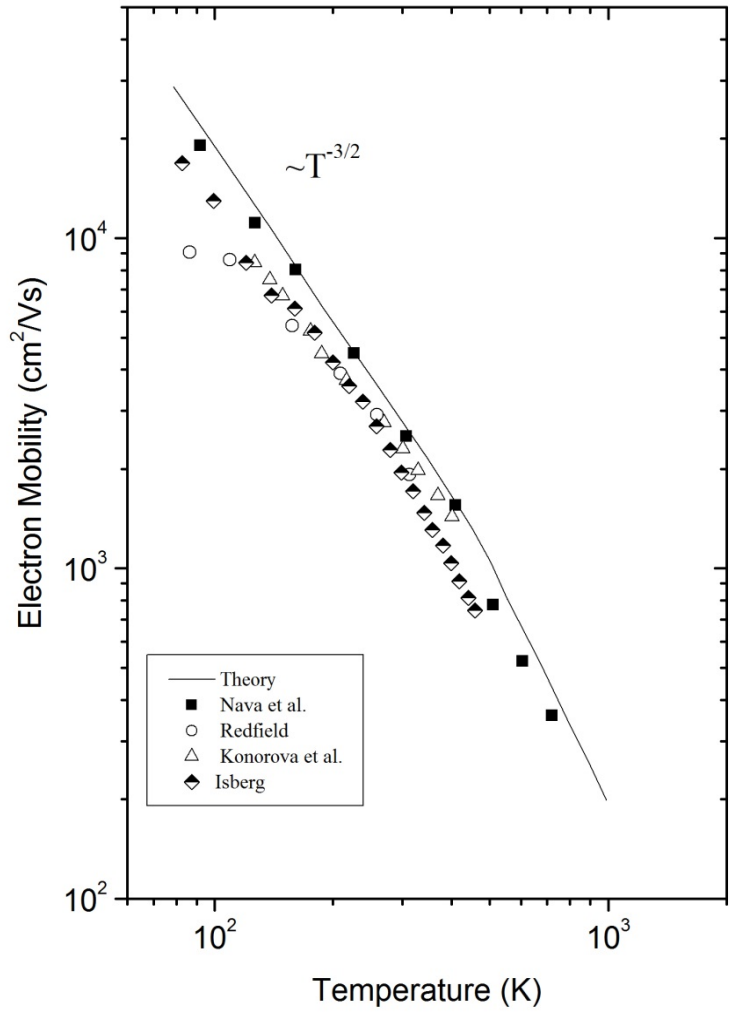

Fig. 5

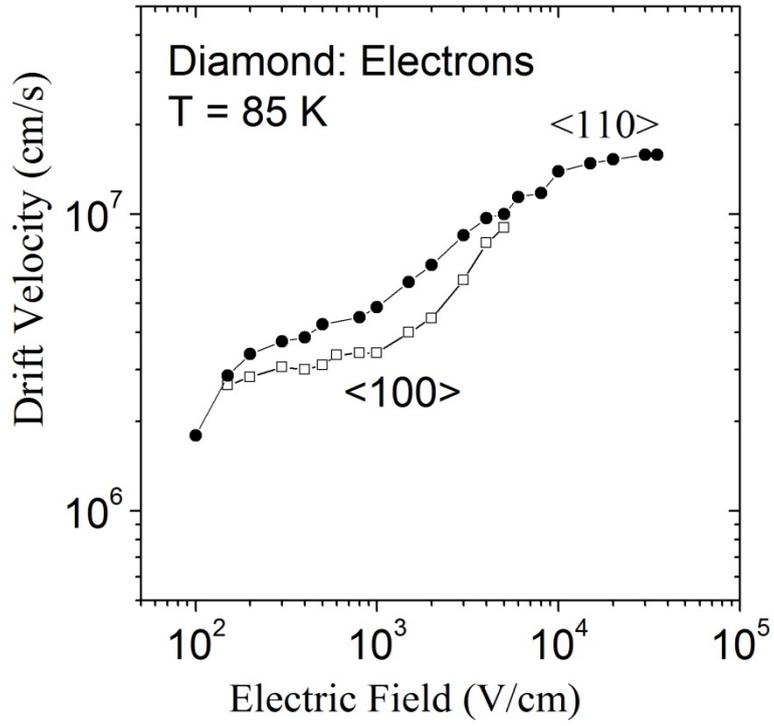

Diamond: Electrons
T = 85 K
<110>
<100>

Fig. 6

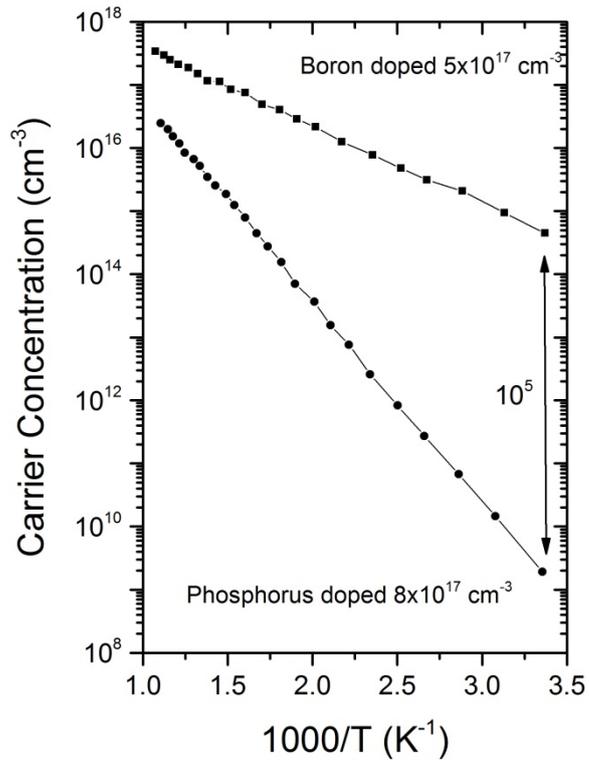

Fig. 7

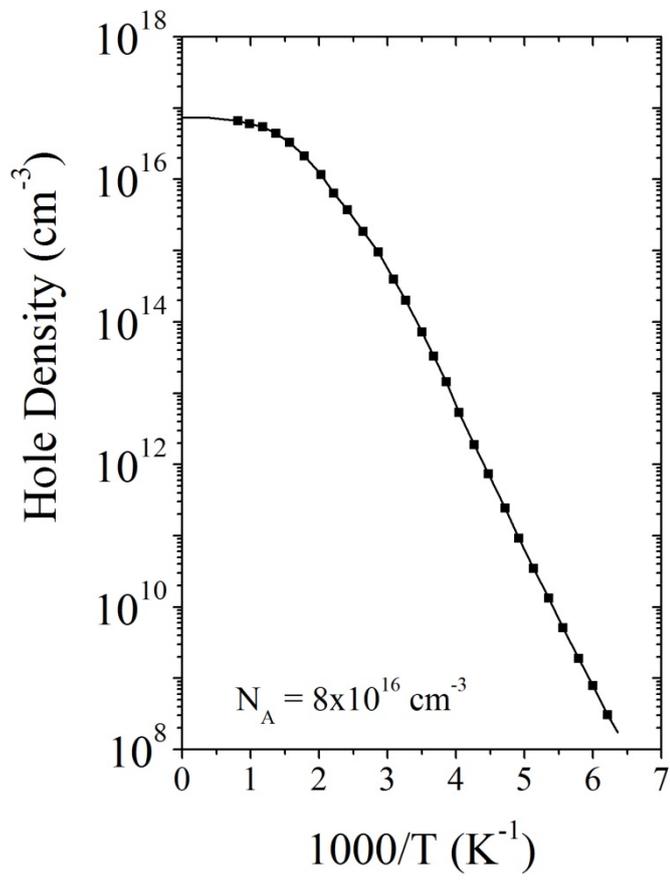

Fig. 8

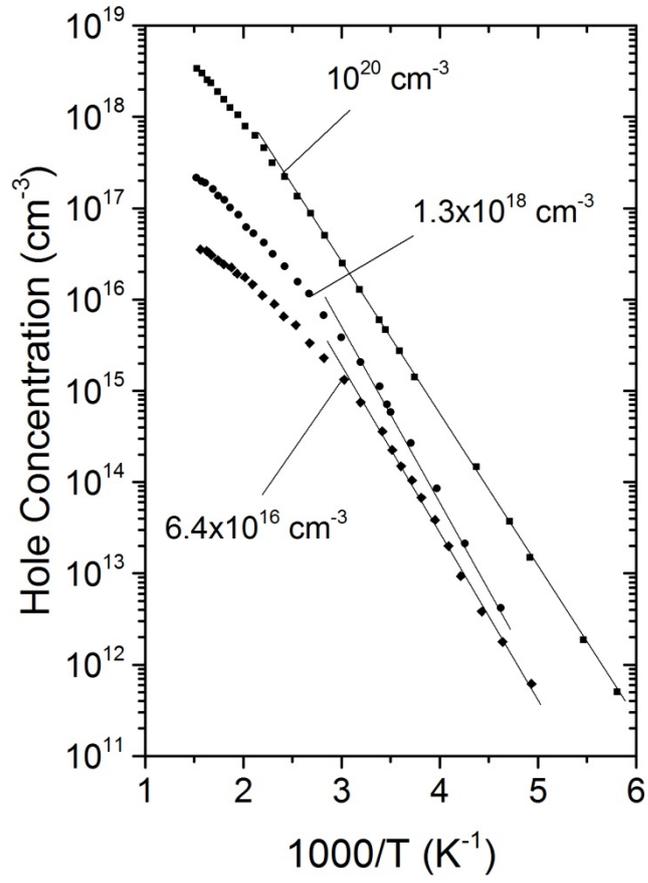

Fig. 9

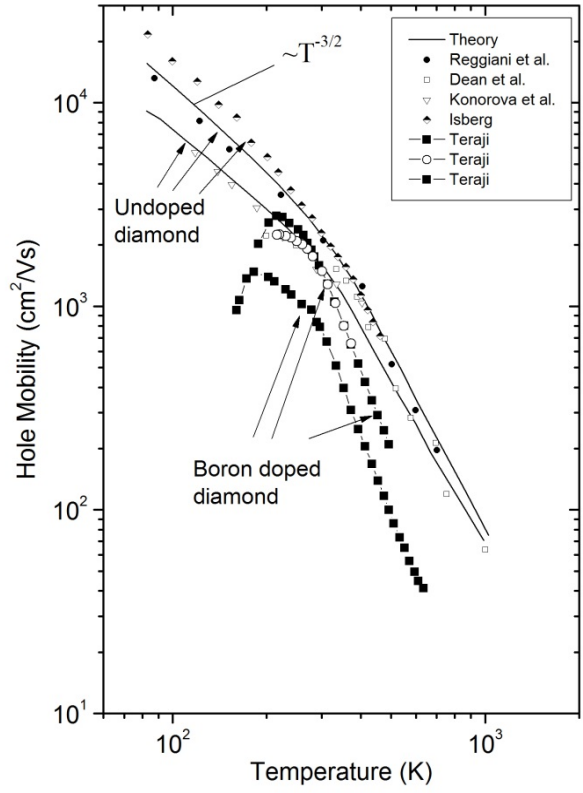

Fig. 10

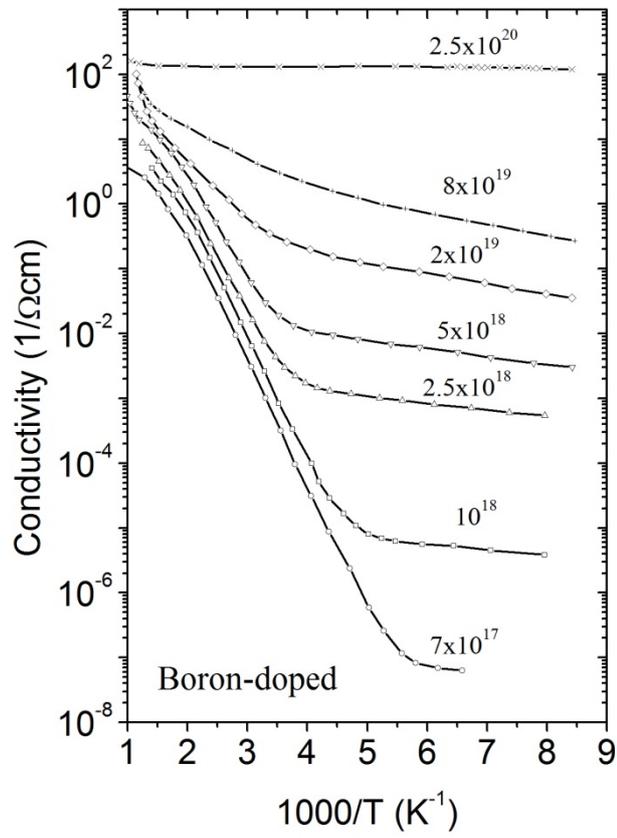

Fig. 11

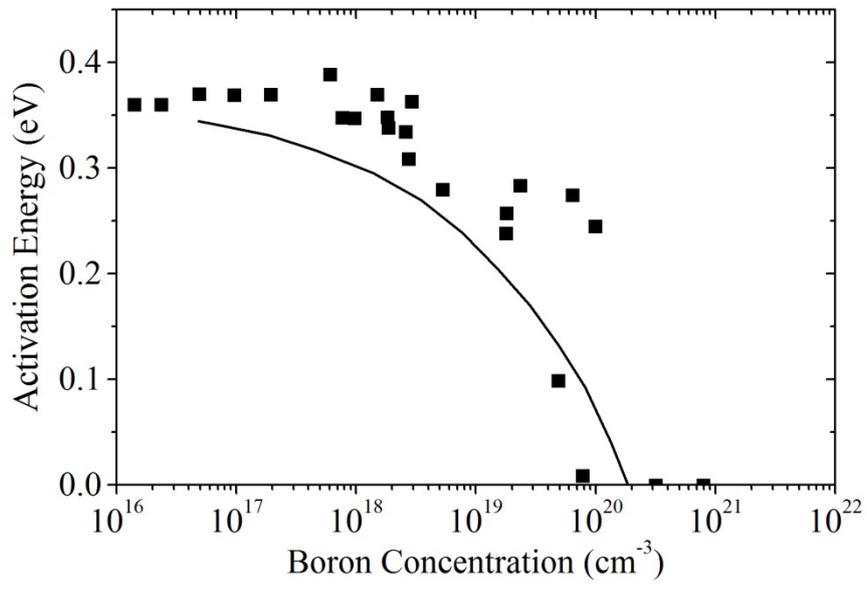

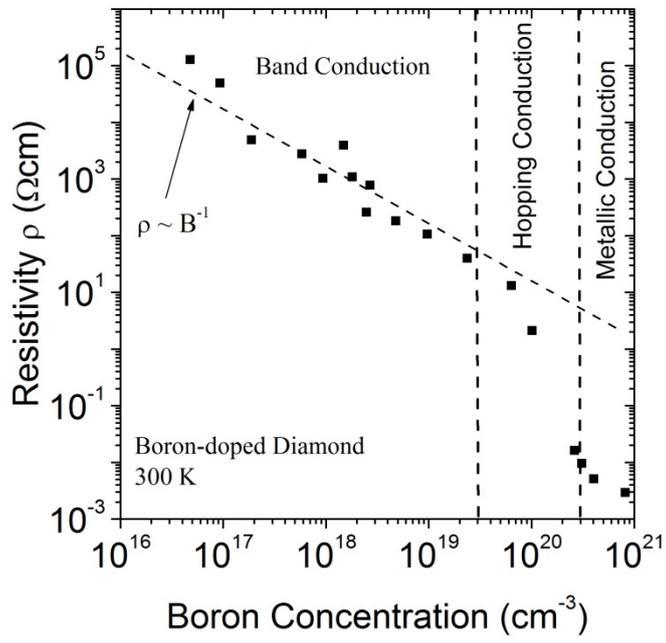

Fig. 13

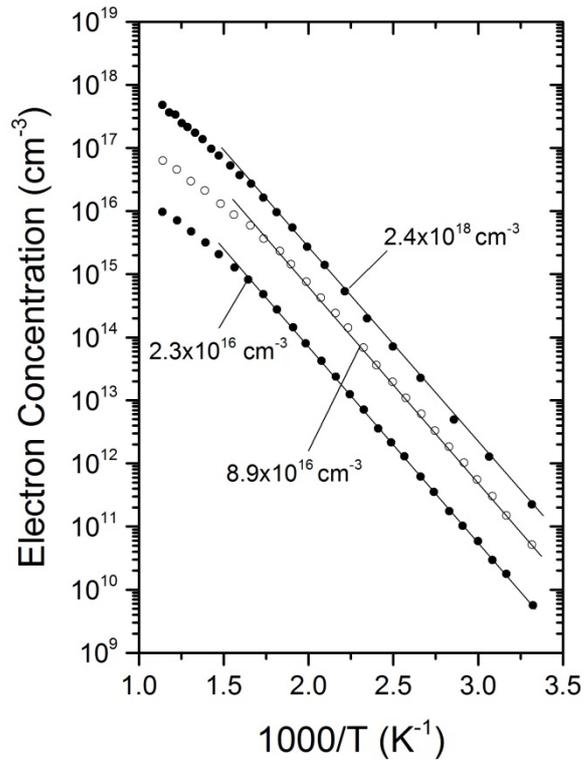

Fig. 14

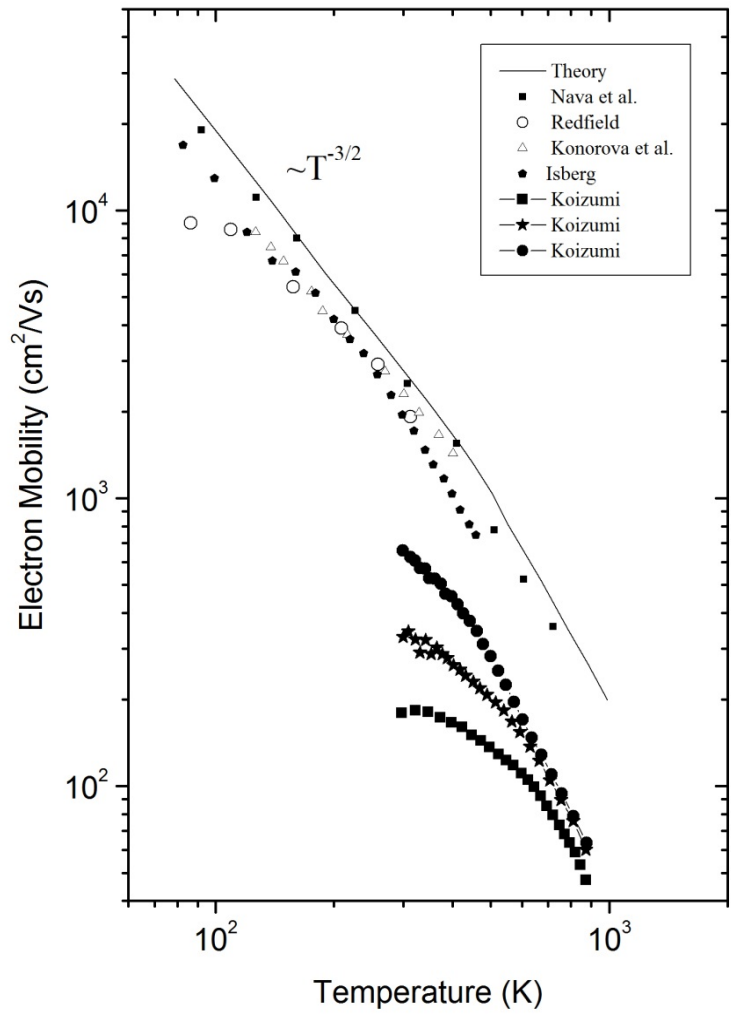

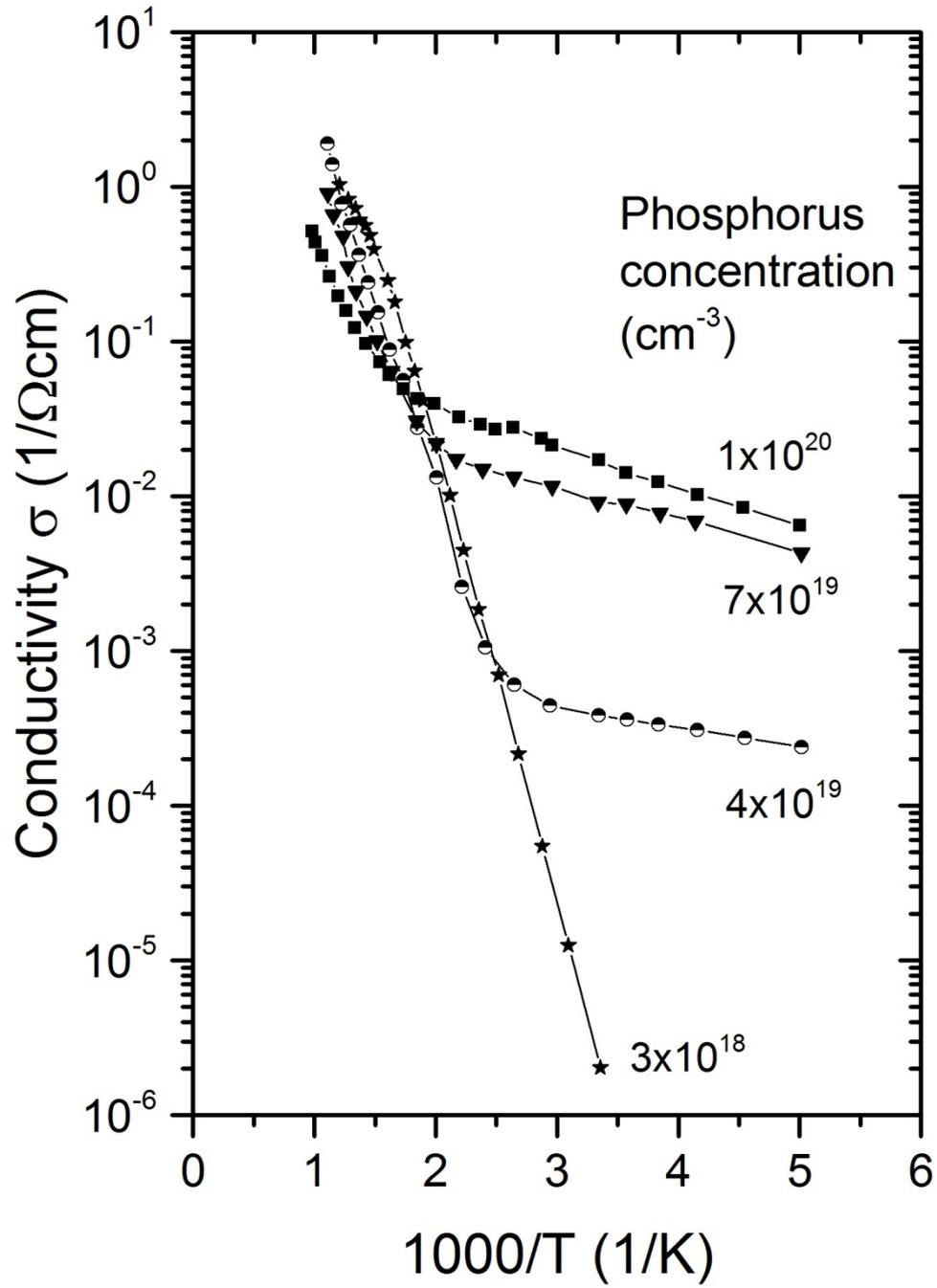

Fig. 16